\documentstyle[12pt,epsf]{article}
\def\largelinestretch{\renewcommand{\baselinestretch}{1.265}}
\largelinestretch\small\normalsize
\title{
\vspace*{-10mm}
\hfill
\parbox{4cm}{\large JINR E2-95-266\\
                    MKPH-T-95-08\\
                    hep-ph/9506406}\\
\vspace*{10mm}
    $\gamma\gamma\to\pi^0\pi^0$ and $\eta\to\pi^0\gamma\gamma$ at $O(p^6)$ \\
in the NJL model}
 \author{
A.A.Bel'kov${}^1$,
A.V.Lanyov${}^1$,
S.Scherer${}^2$\thanks{Supported by Deutsche Forschungsgemeinschaft}
\\
\\ \normalsize
${}^1$
        Particle Physics Laboratory,\\ \normalsize
 Joint Institute for Nuclear Research,\\ \normalsize
        141980 Dubna, Moscow Region, Russia\hfill\\ \normalsize
${}^2$
Institut f\"ur Kernphysik, Johannes Gutenberg--Universit\"at,\\ \normalsize
        D-55099 Mainz, Germany\hfill\\
}
\date{}
\begin{document}
\largelinestretch\normalsize
\thispagestyle{empty}
\begin{titlepage}
\thispagestyle{empty}
\maketitle
\begin{abstract}
   We discuss the processes $\gamma\gamma\to\pi^0\pi^0$ and $\eta\to\pi^0
\gamma\gamma$ at $O(p^6)$ in the momentum expansion.
   The calculation involves tree--level, one--loop and two--loop diagrams
of a chiral effective lagrangian which is obtained by a bosonization of the
NJL model.
   The importance of integrating out meson resonances (reduction) is
pointed out.
   Our final results for the total cross section of
$\gamma\gamma\to\pi^0\pi^0$ are in good agreement with the
experimental data of the Crystal Ball Collaboration.
   For the width of the $\eta\to\pi^0\gamma\gamma$ decay we obtain
the value $0.11$ eV which has to be compared with the experimental value of
$(0.84\pm 0.18)$ eV.
   Alternatively, taking empirical parameters from a vector--meson--dominance
model the prediction for the decay width is $0.35$ eV.
   We present a prediction for the differential decay probability as a
function of $m_{\gamma\gamma}^2/m_\eta^2$.
\end{abstract}
\end{titlepage}


   The theoretical interest in the reaction $\gamma \gamma \to \pi^0 \pi^0$
dates back to the seventies when predictions for the electromagnetic
polarizabilities of the charged as well as the neutral pion were obtained
in the framework of current--algebra techniques \cite{Terentev} and chiral
quantum field theory \cite{Volkov1}.
   These polarizabilities are a signature of the underlying structure of
particles, similar to the electromagnetic root--mean--square radius, and
a large number of different predictions for these parameters has been
obtained in various models (for an overview see, e.g.,
Refs.\ \cite{polarizabilities}).
   The possibility of investigating the $\gamma \gamma \to \pi^0 \pi^0$
amplitude via the $e^+ e^-$--annihilation process as well as the
photoproduction in the Coulomb field of a nucleus was addressed in
Refs.\ \cite{Belkov1}.

   In the meantime, $\gamma \gamma \to \pi^0 \pi^0$ cross section data from
threshold up to the $\rho$--resonance region were provided by the
Crystal Ball Collaboration \cite{crystal-ball}.
   On the theoretical side the framework of Chiral Perturbation Theory
(ChPT) \cite{Weinberg,Gasser1} provides an ideal tool to systematically
study low--energy amplitudes involving Goldstone bosons and their
interactions with external fields, such as the electromagnetic field.
   In Refs.\ \cite{Bijnens1} the amplitude for
$\gamma\gamma\to\pi^0\pi^0$ was calculated to $O(p^4)$ in ChPT,
and the result was found to be given entirely in terms of one--loop
diagrams involving vertices of $O(p^2)$.
   In other words, there are no tree--level diagrams at $O(p^2)$ and
$O(p^4)$ and thus the one--loop diagrams are finite.
   However, the one--loop calculation in ChPT disagrees with the data even
near threshold.
   The inclusion of a Born contribution at $O(p^6)$, obtained either
from quark loops or from vector--meson dominance, results in too small
a contribution to yield agreement with experiment \cite{Bijnens3}.
   On the other hand, the application of dispersive methods leads to a
considerable improvement since they take account of important unitarity
corrections corresponding to rescattering effects of higher order
\cite{dispersion}.
   A full two--loop calculation at $O(p^6)$ within $SU(2)\times SU(2)$ ChPT
was carried out in Ref.\ \cite{Bellucci1}.
   The $O(p^6)$ counterterm contributions were estimated with resonance
saturation and the total result was found to be in good agreement up to an
invariant mass $\sqrt{s}$ of 700 MeV.
   Finally, $\gamma\gamma\rightarrow\pi^0\pi^0$  was also considered in the
context of Generalized ChPT up to one--loop order corresponding to $O(p^5)$
in this counting scheme \cite{Knecht}.

   In the framework of chiral $SU(3)\times SU(3)$ symmetry the decay process
$\eta\rightarrow\pi^0\gamma\gamma$ is closely related to
$\gamma\gamma\rightarrow\pi^0\pi^0$.
   At $O(p^4)$ in ChPT the prediction for the decay width
\cite{Ametller} was found to be two orders of magnitude smaller than the
measured value \cite{PDG}.
   The pion loops are small due to approximate $G$--parity invariance
whereas the kaon loops are suppressed by the large kaon mass in the
propagator.
   A considerable enhancement was obtained with resonance saturation
for some counterterms of higher orders in the momentum expansion.
   In Ref.\ \cite{Ametller} symmetry--breaking terms proportional to
the quark masses were not considered at $O(p^6)$.
   Such counterterms were, however, included in Refs.\ \cite{Bellucci2,Ko}.
   In Ref.\ \cite{Bellucci2} they were estimated in the framework of
an extended NJL model \cite{Bijnens2} whereas in Ref.\ \cite{Ko} the
experimental decay width was used to fit one of the corresponding
coefficients.
   Finally, a complementary approach was used in Ref.\ \cite{Ng} where
the $\eta\rightarrow\pi^0\gamma\gamma$ decay was calculated
in a phenomenological quark model using the quark--box diagram.
   A good agreement with the experimental value for the decay width
was obtained with a constituent quark mass of 300 MeV.

   It is the purpose of this work to present the results of a consistent
calculation of the processes $\gamma \gamma \to \pi^0 \pi^0$ and
$\eta \to \pi^0 \gamma \gamma$ at $O(p^6)$ in the momentum expansion.
   According to Weinberg's power counting scheme \cite{Weinberg} the
calculation involves tree--level, one-- and two--loop diagrams.
   The effective action up to $O(p^6)$ in terms of collective meson
degrees of freedom is obtained by bosonization \cite{bosonization} of
the NJL model \cite{NJL}.
   This effective action, in addition to the pseudoscalar mesons, still
contains scalar, vector and axial--vector degrees of freedom.
   In order to determine the structure coefficients of the effective chiral
lagrangian at $O(p^4)$ \cite{Gasser1} and $O(p^6)$ \cite{Fearing}
one has to integrate out the meson resonances.
   The method of superpropagator regularization \cite{Volkov2} was used
in order to fix the UV divergences which for the first time show up at
$O(p^6)$.

   We start from the generating functional
\begin{equation}
{\cal Z} = \int {\cal D}\Phi\,{\cal D}\Phi^\dagger\,{\cal D}V\,{\cal D}A
           \,\,\mbox{exp}\big[ i{\cal S}(\Phi,\Phi^\dagger ,V,A)\big]\,,
\label{genfunc}
\end{equation}
corresponding to the following action for scalar ($S$),
pseudoscalar ($P$), vector ($V_\mu$) and axial--vector ($A_\mu$)
collective meson fields,
\begin{eqnarray}
{\cal S}(\Phi,\Phi^\dagger ,V,A) &=&
          \int d^4 x
          \bigg [ - {1 \over{4G_1}} \mbox{tr} (\Phi^\dagger\Phi)
           - \, {1 \over{4G_2}} \mbox{tr} (V_\mu V^\mu + A_\mu A^\mu)
\nonumber \\ &&
\quad\quad\quad~ +\mbox{log}(\mbox{det}(i \widehat{\bf D}))\bigg].
\label{action}
\end{eqnarray}
   This action is obtained by first bosonizing the effective action of the
NJL model and then integrating over the quark degrees of freedom.
   In Eq.\ (\ref{action}) $G_{1}$ and $G_{2}$ are parameters which
are fitted to empirical input (see Eqs.\ (\ref{F20}) and (\ref{ZA2})
below for details), $\Phi = S + iP$, and ${\bf \widehat{D}}$
refers to the Dirac operator
\begin{equation}
\label{diracop}
  i{\bf \widehat{D}} =
                       [i(\partial\hspace{-.5em}/ +A\hspace{-.55em}/_R)
                       -(\Phi +m_0)] P_R
                      +[i(\partial\hspace{-.5em}/ +A\hspace{-.55em}/_L)
                       -(\Phi +m_0)^\dagger ]P_L,
\end{equation}
where $m_0$ is the current quark mass matrix,
$P_{R/L}={1 \over 2}(1 \pm \gamma_5)$ are chiral projectors and
$A^{R/L}_\mu = V_\mu \pm A_\mu$.
    The electromagnetic interaction can be included by the
replacement $V_\mu \to  V_\mu +ie{\cal A}_\mu Q$, where $Q$ is the
quark charge matrix, $Q=\mbox{diag}(2/3,-1/3,-1/3)$.  We express $\Phi$
using a nonlinear realization of chiral symmetry,
$$ \Phi = \Omega \,\Sigma \,\Omega, $$
where
$$\Omega(x) = \exp\left(\frac{i}{\sqrt{2}F_0} \varphi (x) \right)\,,$$
\begin{equation}
\varphi =
\left ( \begin{array}{ccc}
\frac{1}{\sqrt{2}}\pi^0+\frac{1}{\sqrt{6}}\eta_8 +
                {1\over\sqrt{3}}\eta_0 &  \pi^+ &  K^+ \\
\pi^- & -\frac{1}{\sqrt{2}}\pi^0+\frac{1}{\sqrt{6}}\eta_8 +
{1\over\sqrt{3}}\eta_0  & K^0 \\
K^- & \bar{K}^0 & -\sqrt{\frac{2}{3}}\eta_8 +{1\over\sqrt{3}}\eta_0
\end{array}
\right )
\end{equation}
represents the pseudoscalar degrees of freedom.
   $F_0$ is the bare $\pi$ decay constant.
   The $3\times 3$ matrix $\Sigma(x)$ contains the scalar fields
and is expanded around its vacuum expectation value $\mu$,
\begin{equation}
\Sigma(x)=\mu+\sigma(x).
\label{Sigma}
\end{equation}
   The constituent quark mass $\mu$ is the solution of the gap equation.

   For the processes under consideration, up to and including $O(p^6)$,
only the even--intrinsic--parity sector of the chiral lagrangian
is required \cite{Ametller}.
   This sector is obtained from the modulus of the logarithm of the
quark determinant and can be calculated using the heat--kernel technique
with proper--time regularization \cite{propertime,Belkov2}.
   This method has been used in Ref.\ \cite{Belkov3} to obtain a prediction
for the structure coefficients of the general effective lagrangian
of $O(p^4)$ and $O(p^6)$, respectively \cite{Gasser1,Fearing}.
   The result of Ref.\ \cite{Belkov3} explicitly contains, apart from the
pseudoscalar Goldstone bosons, scalar, vector and axial--vector
resonances as {\em dynamical} degrees of freedom.
   However, in order to avoid double counting when calculating processes
involving Goldstone bosons and photons, one has to integrate
out (reduce) these resonances in the generating functional of
Eq.~(\ref{genfunc}) and thus one effectively takes resonance--exchange
contributions into account.
   As a consequence of this procedure the structure coefficients of
pseudoscalar low--energy interactions will be strongly modified
\cite{Bijnens2,reduction,Belkov4}.

   In order to perform the integration over the scalar, vector and
axial--vector fields in Eq.\ (\ref{genfunc}) we made use of the fact that
the modulus of the quark determinant in Eq.\ (\ref{action}) is invariant
under local chiral transformations of the fields \cite{Belkov4,Reinhardt}.
   This allows us, with a specific choice for the chiral transformation
(unitary gauge), to eliminate the pseudoscalar fields from the Dirac
operator, Eq.\ (\ref{diracop}).
   At the same time, introducing $\Phi'= \Phi -m_0$ and renaming
$\Phi'\to\Phi$ generates the mass term for the pseudoscalars from the
Gaussian part of Eq.\ (\ref{action}).
   Furthermore, interactions between the pseudoscalar degrees of freedom
and the transformed vector and axial--vector fields are generated in
the Gaussian part.
   The masses of the scalar, vector and axial--vector mesons are
sufficiently large in comparison with the Goldstone boson masses, and
thus it is possible to integrate out the meson resonances using their
respective equations of motion in the static limit.
   These equations result from a variation of the effective action of
Eq.\ (\ref{action}) by neglecting terms of $O(p^4)$ and higher in the
logarithm of the quark determinant.
   The remaining part of the action then is quadratic in the resonances,
in particular, there are no terms containing field strength tensors.

    The invariant amplitude
${\cal M}=i\epsilon^\mu_1 \epsilon^\nu_2 T_{\mu \nu}$ of the process
$\gamma (q_1) \gamma (q_2) \to a(p_1) b(p_2)$ can be expressed in terms of
two functions $A$ and $B$ as
\begin{eqnarray}
T_{\mu\nu}^{\gamma \gamma \to ab} &=&
  A(s,\nu) \left( \frac{s}{2} g_{\mu\nu} -q_{2\mu}q_{1\nu}\right)
\nonumber \\ &&
 +B(s,\nu)\Big[ 2s\Delta_\mu \Delta_\nu
                -\Big( \nu^2 - \big(m^2_b - m^2_a \big)^2 \Big)
                 g_{\mu \nu}
\nonumber \\ &&
  +2\Big( \big( \nu + m^2_b - m^2_a \big)q_{2\mu}\Delta_\nu
         -\big( \nu + m^2_a - m^2_b \big)\Delta_\mu q_{1\nu}
    \Big) \Big],
\label{ampdef1}
\end{eqnarray}
where $s=(q_1+q_2)^2$, $\nu = 2p_1\!\cdot\!(q_2-q_1)$, and
$\Delta_{\mu}=(p_1-p_2)_{\mu}$.
   The amplitude for the process
$a(p_1)\to b(p_2)\gamma (q_1) \gamma (q_2)$ can be obtained from Eq.\
(\ref{ampdef1}) using crossing symmetry, namely, by performing the
replacement $q_i \to -q_i$ and $p_1 \to -p_1$.
   However, for the decay channel
$\eta(k) \to \pi^0(p) \gamma(q_1) \gamma(q_2)$, it turns out to be more
convenient to use the parameterization
\begin{eqnarray}
  T^{(\eta \to \pi^0 \gamma \gamma)}_{\mu\nu} &=&
  {\cal{A}}(x_1,x_2)\big[g_{\mu\nu}(q_1\cdot q_2)-q_{1\nu}q_{2\mu}\big]
\nonumber \\ &&
 +{\cal{B}}(x_1,x_2)\bigg[ m^2_{\eta} x_1 x_2 g_{\mu\nu}
       +\frac{(q_1\cdot q_2)}{m^2_{\eta}}k_{\mu}k_{\nu}
       -x_1 q_{2\mu} k_{\nu}-x_2 k_{\mu} q_{1\nu}\bigg]\,,
\label{ampdef2}
\end{eqnarray}
where $x_i=(k\cdot q_i)/m^2_{\eta}$.

   The prediction for the amplitudes of Eqs.\ (\ref{ampdef1}) and
(\ref{ampdef2}) will involve the structure coefficients $L_i$ of
the Gasser--Leutwyler lagrangian in one--loop diagrams at $O(p^6)$
as well as new coefficients $d_i$ from Born diagrams at $O(p^6)$.
   It is straightforward to obtain the effective lagrangian at $O(p^6)$
contributing to the processes under consideration from the most general
representation of Ref.\ \cite{Fearing},
\footnote{Note that there are different conventions for the definition
of the coefficients $d_i$.}
\begin{eqnarray}
{\cal L}_6&=& \frac{8}{F_0^2}\left[
 d_1 {\cal F}_{\mu\alpha} {\cal F}^{\mu\beta}
 \mbox{tr}\left(\partial^\alpha U_0
\partial_\beta U_0^\dagger Q^2\right)
+d_2 {\cal F}_{\mu\nu} {\cal F}^{\mu\nu}\mbox{tr}
 \left(\partial_\alpha U_0 \partial^\alpha U_0^\dagger Q^2\right) \right.
\nonumber\\&&
+d_3 {\cal F}_{\mu\nu} {\cal F}^{\mu\nu} \mbox{tr}
 \left(\chi (U_0+U_0^\dagger )Q^2 \right)
+d_4 {\cal F}_{\mu\nu} {\cal F}^{\mu\nu} \mbox{tr}
 (Q^2)\mbox{tr}\left(\chi (U_0+U_0^\dagger)\right)
\nonumber\\&&
+d_5 {\cal F}_{\mu\alpha} {\cal F}^{\mu\beta} \mbox{tr}\left(Q^2\right)
 \mbox{tr}\left(\partial^\alpha U_0 \partial_\beta U_0^\dagger \right)
+d_6 {\cal F}_{\mu\nu} {\cal F}^{\mu\nu} \mbox{tr}\left(Q^2\right)
 \mbox{tr}\left(\partial_\alpha U_0 \partial^\alpha U_0^\dagger \right)
\nonumber\\&&
+ d_7 {\cal F}_{\mu\alpha} {\cal F}^{\mu\beta}
\mbox{tr}\left(\partial^\alpha U_0 U_0^\dagger Q\right)
\mbox{tr}\left(\partial_\beta U_0 U_0^\dagger Q\right)
\nonumber\\&&
\left.
+d_8 {\cal F}_{\mu\nu} {\cal F}^{\mu\nu} \mbox{tr}
  \left(\partial_\alpha U_0 U_0^\dagger Q\right)
  \mbox{tr}\left(\partial^\alpha U_0 U_0^\dagger Q \right)
\right].
\label{genlag}
\end{eqnarray}
   In Eq.\ (\ref{genlag}), ${\cal F}_{\mu\nu}=\partial_\mu {\cal A}_\nu-
\partial_\nu {\cal A}_\mu$ is the ordinary electromagnetic field
strength tensor,
   $$U_0=\exp(i\frac{\sqrt{2}\varphi_0}{F_0}),$$
$$
\varphi_0=\mbox{diag}\left(\frac{\pi^0}{\sqrt{2}}+\frac{\eta_8}{\sqrt{6}}
                                +\frac{1}{\sqrt3}\eta_0,\,
-\frac{\pi^0}{\sqrt{2}}+\frac{\eta_8}{\sqrt{6}}+\frac{1}{\sqrt3}\eta_0,\,
-\sqrt{\frac{2}{3}}\eta_8+\frac{1}{\sqrt3}\eta_0\right),$$
and $\chi \equiv \mbox{diag}(\chi^2_u,\chi^2_d,\chi^2_s)
=-2m_0<\!\!\bar{q}q\!\!>F^{-2}_0$ is the mass
matrix, where $<\!\!\bar{q}q\!\!>$ is the quark condensate.
   Previous calculations considered the counterterms of Eq.\ (\ref{genlag})
with various degrees of approximation.
   In Ref.\ \cite{Ametller} only single--trace terms in the chiral limit
were taken into account.
   In Refs.\ \cite{Bellucci1,Bellucci2} the chiral symmetry breaking term
proportional to $d_3$ was included and Ref.\ \cite{Ko}
also took $d_4$ into account.
   The double--trace terms proportional to $d_4$ -- $d_8$ typically do
not appear in  effective lagrangians derived from the bosonization of
NJL type quark models.

   In the NJL model only the structure constants $d_1$, $d_2$, $d_3$
contribute to the Born amplitudes of the processes
$\gamma\gamma\to\pi^0\pi^0$ and $\eta\to\pi^0\gamma\gamma$ at $O(p^6)$,
respectively,
\begin{eqnarray}
  A^{B(p^6)} &=& \frac{64e^2}{9F^4_0}\bigg[
       \frac{5}{16}d_1s + \frac{5}{2}d_2(s-2m^2_{\pi})
      +d_3(4\chi^2_u + \chi^2_d) \bigg]\,,
\nonumber \\
  B^{B(p^6)} &=& -\frac{10e^2}{9F^4_0} d_1\,,
\label{ampggpp}
\end{eqnarray}
and
\begin{eqnarray}
  {\cal{A}}^{B(p^6)} &=& \frac{8e^2}{3\sqrt{3}F^4_0}
        C_{\theta}\bigg\{
        2(d_1+4d_2)m^2_{\eta}(x_1+x_2)
       -\frac{8}{3}\bigg[3d_2m^2_{\eta}+d_3(-4\chi^2_u+\chi^2_d)\bigg]
\nonumber \\ &&
       +\frac{\chi^2_u-\chi^2_d}{6(m^2_{\eta}-m^2_{\pi})}\bigg[
        (d_1+4d_2)m^2_{\eta}(x_1+x_2)
       -4d_2m^2_{\eta}+4d_3(4\chi^2_u+\chi^2_d) \bigg]
\nonumber \\ &&
       -\frac{1}{3} (\chi^2_u-\chi^2_d)\Theta_1 \bigg[~
        (d_1+4d_2)m^2_{\eta}(x_1+x_2)
\nonumber \\ &&~~~~~~~~~~~~~~~~~~~~
       -\frac{4}{3}\bigg( d_2m^2_{\eta}
                  -d_3(4\chi^2_u+\chi^2_d+4\chi^2_s)\bigg)
       \bigg]\bigg\}\,,
\nonumber \\
  {\cal{B}}^{B(p^6)} &=& -\frac{16e^2}{3\sqrt{3}F^4_0}
        C_{\theta}\bigg[ 2
       +\frac{5}{3}\frac{\chi^2_u-\chi^2_d}{m^2_{\eta}-m^2_{\pi}}
       +\frac{1}{3}(\chi^2_u-\chi^2_d)\Theta_1 \bigg]m^2_{\eta}d_1\,.
\label{ampeta}
\end{eqnarray}
   In Eqs.\ (\ref{ampeta})
$C_{\theta}=\mbox{cos}\theta-\sqrt{2}\mbox{sin}\theta$, where
$\theta=-19^o$ is the $\eta\!-\!\eta'$ mixing angle,
$$\eta_8 =  \eta \cos \theta + \eta' \sin \theta\,, \quad
  \eta_0 = -\eta \sin \theta + \eta' \cos \theta\,,
$$
and furthermore we have introduced
\begin{eqnarray*}
\Theta_1 &=& \frac{(\mbox{cos}\theta-\sqrt{2}\mbox{sin}\theta )^2}
                {m^2_{\eta}-m^2_{\pi}}
          +\frac{(\mbox{sin}\theta+\sqrt{2}\mbox{cos}\theta )^2}
                {m^2_{\eta^{'}}-m^2_{\pi} }\,.
\end{eqnarray*}
   Note that the $\eta$ decay amplitudes of Eqs.\ (\ref{ampeta}) also include
contributions of the pole diagrams with $\pi^0\!-\!\eta$ and
$\pi^0\!-\!\eta'$ transitions.

   We now turn to the determination of the structure coefficients
within the framework of the NJL model.
   It is a well--known fact that the elimination of the resonance degrees of
freedom gives rise to a substantial modification of the structure constants.
   At $O(p^2)$ such a reduction leads to a redefinition of the decay
constant $F_0$ and the mass matrix $\chi$.
   To be specific, the identification of the decay constant before and
after reduction is given by
\begin{equation}
F^2_0 =  \frac{N_c \mu^2 y}{4\pi^2}\quad \longrightarrow\quad
F^2_0 =  Z^2_A \frac{N_c \mu^2 y}{4\pi^2}\,,
\label{F20}
\end{equation}
respectively, and similarly
\footnote{Using the gap equation it can be shown that both
          expressions for $\chi$ in Eqs.\ (\ref{chi}) are equivalent
          for $\mu^2/\Lambda^2\ll1$.}
for $\chi$
\begin{equation}
\chi = -2m_0\mu \bigg( 1-
        \frac{\Lambda^2}{y\mu^2} \mbox{e}^{-\mu^2/\Lambda^2}
                \bigg)
\quad\longrightarrow\quad
\chi = \frac{m_0\mu}{G_1F_0^2}\,,
\label{chi}
\end{equation}
where $y=\Gamma\big(0,\mu^2/\Lambda^2\big)$, $\mu$ is the average
constituent quark mass,  $\Lambda$ is the intrinsic
cutoff parameter, and
\begin {equation}
Z_A^{-2} = 1 +\bigg(\frac{g^0_V}{m^0_V}\bigg)^2
              \frac{N_c\mu^2y}{4\pi^2}\,,\quad
\bigg(\frac{m^0_V}{g^0_V}\bigg)^2=\frac{1}{4G_2}.
\label{ZA2}
\end{equation}
   The incomplete gamma function is defined as $\Gamma(n,x)=
\int_x^\infty dt\, e^{-t} t^{n-1}$.
   In Eq.\ (\ref{ZA2}) we have introduced
$$
g^0_V = \bigg[ \frac{N_c}{48 \pi^2}\big(2y-1\big)\bigg]^{-1/2}\,,
\quad
(m^0_V)^2 = m_{\rho}^2 (1+\tilde{\gamma})\,,
\quad
\tilde{\gamma}=\frac{N_c(g^0_V)^2}{48\pi^2}\,.
$$
   The parameter $Z_A^2$ of Eq.\ (\ref{ZA2}) corresponds to the
$\pi - A_1$ mixing factor and has the phenomenological value
$$
  Z^2_A = \frac{m^2_{\rho}}{m^2_{A_1}}
          \frac{1+\tilde{\gamma}}{1-\tilde{\gamma}} \approx 0.62,
$$
where we used the following empirical input, $m_{\rho} = 770$ MeV,
$m_{A_1} = 1260$ MeV, and $g_V = g_{\rho \pi \pi} = 6.3$.
    On the other hand, with the special choice $Z^2_A = 1/2$,
Eqs.~(\ref{ZA2}) and (\ref{F20}) reproduce the well--known
Kawarabayashi--Suzuki relation, $m^2_{\rho} = 2 g^2_VF^2_0$.

   A full calculation of the $\pi$ and $K$ decay constants at $O(p^4)$
allows to fix the parameters $y$ and
$x=-\mu F^2_0/( 2\!<\!\!\overline{q}q\!\!>\!)$ for given values of $Z_A^2$
and $\mu$, by identifying the decay constants with their empirical values.
   In the following we will use $Z_A^2=0.62$ and $\mu =
   265\mbox{~MeV}$,
from which we obtain $y = 2.4$ and $x = 0.10$.
   These values correspond to $F_0 = 90 \mbox{~MeV}$ and
$<\!\!\overline{q}q\!\!>^{1/3} = -220 \mbox{~MeV}$.

   At $O(p^4)$ the reduction of the resonances \cite{Belkov4} leads to the
following modification of the structure coefficients of the lagrangian
introduced by Gasser and Leutwyler \cite{Gasser1}
$\big(L_i = \frac{N_c}{16\pi^2}l_i\big)$,
\begin{eqnarray}
l_1 = \frac{1}{24}\,, &&
l^{red}_1 = \frac{1}{24}\bigg[ Z^8_A
          +2(Z^4_A-1)\bigg(\frac{1}{4}y(Z^4_A-1)-Z^4_A\bigg)\bigg]
          = 1.08\, l_1\,;
\nonumber \\
l_2 = 2 l_1\,, &&
l^{red}_2 = 2l^{red}_1\,;
\nonumber \\
l_3 = -\frac{1}{6}\,, &&
l^{red}_3 = -\frac{1}{6}\bigg[ Z^8_A
          +3(Z^4_A-1)\bigg(\frac{1}{4}y(Z^4_A-1)-Z^4_A\bigg)\bigg]
          =1.54\, l_3\,;
\nonumber \\
l_4 = 0\,, && l^{red}_4=0\,;
\nonumber \\
l_5 = x(y-1)\,, &&
l^{red}_5 = (y-1)\frac{1}{4}Z_A^6=0.60\, l_5\,;
\nonumber \\
l_6 = 0\,, && l^{red}_6=0\,;
\nonumber \\
l_7 = -\frac{1}{6}\bigg(xy-\frac{1}{12}\bigg)\,, && l^{red}_7=0\,;
\nonumber\\
l_8 = \bigg(\frac{1}{2}-x\bigg) xy-\frac{1}{24}\,, &&
l^{red}_8 = \frac{y}{16}Z_A^4=1.07\, l_8\,;
\nonumber\\
l_9 = \frac{1}{3}\,, &&
l^{red}_9 = \frac{1}{3}\bigg( Z^4_A-\frac{1}{2}y(Z^4_A-1)\bigg)
          = 1.12\, l_9\,;
\nonumber \\
l_{10} = -\frac{1}{6}\,, &&
l^{red}_{10} = -\frac{1}{6}\bigg(Z^4_A-y(Z^4_A-1)\bigg)= 1.86\, l_{10}\,.
\label{lred}
\end{eqnarray}
   In order to obtain the expressions for the reduced coefficients of
Eq.\ (\ref{lred}), the static equations of motion of the scalar, vector and
axial--vector resonances  have been applied.
   In such an approach scalar resonances can only modify $l_5$ and $l_8$.
   Note that the above results are in agreement with those obtained in
Ref.\ \cite{Bijnens2} except for $l_3^{red}$ and $l_8^{red}$
(see Sect 5.5 of Ref.\ \cite{Bijnens2}).
   The disagreement originates in a different procedure of
integrating out the scalar resonances.
   We will come back to this point below when discussing higher--order
corrections to the static equations of motion.

   We will now discuss those structure constants $d_i$ at $O(p^6)$ which do
not vanish in the NJL model.
   Before reduction we obtain
\begin{eqnarray}
&&d_1 =-\frac{N_c}{16\pi^2}\frac{F^2_0}{\mu^2}\frac{1}{24}
      =-9.13\times 10^{-5}\,,\quad
  d_2 = \frac{N_c}{16\pi^2}\frac{F^2_0}{\mu^2}\frac{1}{48}
      = 4.57\times 10^{-5}\,,\nonumber\\
&&d_3 =\frac{N_c}{16\pi^2}\frac{F^2_0}{\mu^2}\frac{1}{12}x
      = 1.83\times 10^{-5}\,.
\label{di}
\end{eqnarray}
   The first two constants coincide with the results of Ref.\ \cite{Bijnens3}.
The reduction of meson resonances in the framework of applying the
static equations of motion generates the following modifications
\begin{eqnarray}
&&d_1^{red} =-\frac{N_c}{16\pi^2}\frac{F^2_0}{\mu^2}\frac{1}{24}Z^4_A
            =-3.51\times 10^{-5}\,,\quad
  d_2^{red} = \frac{N_c}{16\pi^2}\frac{F^2_0}{\mu^2}\frac{1}{48}Z^4_A
            = 1.76\times 10^{-5}\,,\nonumber\\
&&d_3^{red} =\frac{N_c}{16\pi^2}\frac{F^2_0}{\mu^2}\frac{1}{96}Z^2_A
            =1.42\times 10^{-5}\,.
\label{dired}
\end{eqnarray}
   In this context we note that the modification of the first two
structure coefficients results from the application of the equation of motion
to vector and axial--vector resonances.
   This change amounts to a multiplication of the original coefficients
$d_1$ and $d_2$ of Eq.\ (\ref{di}) by a factor $Z_A^4$.
   The situation for $d_3$ is qualitatively different. In this case
the application of the equation of motion to the {\em scalar} resonances
modifies this coefficient.
   Let us compare our results for $d_i^{red}$ with those of
Ref.\ \cite{Bellucci2}.
   We agree for the coefficients $d_1^{red}$ and $d_3^{red}$
but differ with respect to $d_2^{red}$.
   In order to understand this discrepancy we note that two different
techniques were used to eliminate the resonances.
   In the treatment of scalar resonances the method of Ref.\
\cite{Bellucci2} involves operators with derivatives which are beyond
the scope of our treatment using the static equation of motion.
   A comparison with Eqs.\ (23), (32) and (38) of Ref.\ \cite{Bellucci2}
shows that such operators are the origin for the difference in $d_2^{red}$.
   However, there is another interesting observation.
   Even though our final expression for $d_3^{red}$ is the same as
Eq.\ (40) of Ref.\ \cite{Bellucci2} our result originates entirely from the
reduction of scalar resonances whereas in Ref.\ \cite{Bellucci2}
it is the sum of a scalar resonance contribution (see Eq.\ (39))
and a quark--loop contribution (see Eq.\ (23)) for which we have no
analogue.

   Finally, we have also investigated in our approach those results of
Ref.\ \cite{Bellucci2} which correspond to the inclusion of operators
containing derivatives when integrating out the scalar resonance.
   To this end, after a unitary gauge transformation of the modulus of
the quark determinant, one has to keep also higher--order terms in the
effective action of Eq.\ (\ref{action}) which are linear in the
scalar field $\sigma (x)$ and which contain the coupling to vector,
axial--vector fields and field strength tensors.
   Such higher--order terms lead to a modification of the static equation of
motion for the scalar resonances and thus give an {\em additional} contribution
to the structure coefficients $l^{red}_3$ and $d^{red}_2$,
\begin{equation}
l_3^{red(h.o.)}=\frac{1}{4}\frac{(y-1)^2}{y}Z^8_A
               = -0.18 l_3\,,
\label{l3higher}
\end{equation}
\begin{equation}
d_2^{red(h.o.)}=\frac{N_c}{16\pi^2}\frac{F^2_0}{\mu^2}\frac{1}{48}
                                   \frac{y-1}{y}Z^4_A
               = 1.02\times 10^{-5}\,,
\label{d2higher}
\end{equation}
   which agree with Eq.\ (155) of Ref.\ \cite{Bijnens2} and Eq.\ (38)
of Ref.\ \cite{Bellucci2}, respectively.
   The total result for the coefficients $l^{red}_3$ and $d^{red}_2$ after
reduction of the vector, axial--vector and scalar degrees of freedom then is
the sum of the contributions of Eqs.\ (\ref{lred}) and (\ref{l3higher}) and
(\ref{dired}) and (\ref{d2higher}), respectively.
   It is worth noting that the considered higher--order terms also modify
the static equation of motion of axial--vector resonances.
However, this modification does not lead to any new contributions for
either the structure coefficients $L_i$ or $d_i$.

\begin{figure}[t]
Table 1. Modification of the coefficients $a_1$, $a_2$ and $b$ of
Eq.\ (\ref{abcoeff}) due to the reduction of meson resonances.
${\cal N}=N_c(4\pi F_0/\mu)^2=54.6$, $Z_A^2=0.62$.
\begin{center}
\scriptsize
\renewcommand{\arraystretch}{1.4}
\begin{tabular}{|c|c|c|c|c|c|} \hline
      &Without  &\multicolumn{4}{|c|}{Reduction of resonances}
\\ \cline{3-6}
Coeff.&reduction&$V_\mu$-- and $A_\mu$--fields
      &\multicolumn{2}{|c|}{$\sigma$--field}&Sum
\\ \cline{4-5}
      & &in static approx.&Static approx.
        &Higher--order correct.&
\\ \hline
$a_1$ & $\frac{20}{27}(13x-1){\cal N}=12.1$
      & $-\frac{20}{27}Z^4_A{\cal N}=-15.6$
      & $\frac{10}{27}Z^2_A{\cal N}=12.5$
      & $-\frac{20}{27}Z^2_A\big(1-\frac{1}{y}\big){\cal N}=-9.0$
      & -12.1
\\ \hline
$a_2$ & $\frac{5}{18}{\cal N}=15.2$
      & $\frac{5}{18}Z^4_A{\cal N}=5.8$
      & $ 0$
      & $\frac{10}{27}Z^4_A\big(1-\frac{1}{y}\big){\cal N}=4.5$
      & 10.3
\\ \hline
$b$   & $\frac{5}{108}{\cal N}=2.53$
      & $\frac{5}{108}Z^4_A{\cal N}=0.97$
      & $0$ & $0$ &0.97
\\ \hline
\end{tabular}
\renewcommand{\arraystretch}{1}
\end{center}
\end{figure}

   For the purpose of comparing our numerical results for $\gamma\gamma\to
\pi^0\pi^0$ with those of Refs.\ \cite{Bellucci1,Bellucci2}, it is convenient
to introduce the following parameterization \cite{Bellucci1} of the Born
contribution at $O(p^6)$ for the amplitudes $A$ and $B$ of
Eq.\ (\ref{ampdef1}),
\begin{equation}
\label{abcoeff}
A_6=\frac{a_1 m^2_\pi + a_2 s}{(16 \pi^2 F^2_0)^2},\quad
B_6=\frac{b}{(16 \pi^2 F^2_0)^2}.
\end{equation}
The coefficients $a_1$, $a_2$ and $b$ are related to $d_1$, $d_2$ and $d_3$ by
\begin{eqnarray*}
a_1 = (4\pi )^4\frac{10}{9}32(d_3-d_2)\,,\quad
a_2 = (4\pi )^4\frac{10}{9}2(d_1+8d_2)\,,\quad
b   =-(4\pi )^4\frac{10}{9}d_1\,.
\end{eqnarray*}
   Our results for $a_i$ and $b$ are summarized in Table 1.
   Clearly, the reduction of the resonances leads to a large modification of
the coefficients.
   However, one has to keep in mind that the effective action after the
reduction describes the interaction of only pseudoscalars and photons.
   Thus the modified coefficients should not be treated as {\em additional}
corrections to the nonreduced coefficients of Eq.\ (\ref{di}).
   A summation of quark--loop contributions and resonance--exchange
contributions to the structure coefficients as in Table 1 of
Ref.\ \cite{Bellucci2}, in our opinion, leads to double counting.

   Before comparing our values of the $O(p^6)$ structure coefficients
with those of Ref.\ \cite{Bellucci1} we provide a prescription for relating
results in different renormalization schemes.
   In our approach UV divergences, resulting from meson loops at $O(p^6)$,
were separated using the superpropagator regularization method
\cite{Volkov2} which is particularly well--suited for the treatment of
loops in nonlinear chiral theories.
   The result is equivalent to the dimensional regularization technique
used in Ref.\ \cite{Bellucci1}, the difference being that the scale parameter
$\mu$ is no longer arbitrary but fixed by the inherent scale of the chiral
theory, namely, $\tilde{\mu} = 4\pi \, F_0$.
   In order to compare the two methods the UV divergences have to be replaced
by a finite term using the substitution
$$
(C - 1/\varepsilon) \quad \longrightarrow \quad C_{SP}
=2C+1+\frac12\,\left[{d \over dz}
                     \left(\log \Gamma^{-2}(2z+2)
                     \right)
               \right]_{z=0} + \beta \pi
=-1+4C + \beta \pi\,,
$$
   where $C=0.577$ is Euler's constant, $\varepsilon=(4-D)/2$,
and $\beta$ is an arbitrary constant resulting from the
Sommerfeld--Watson integral representation of the superpropagator.
   The splitting of the decay constants $F_\pi$ and $F_K$ is used at $O(p^4)$
to fix $C_{SP} \approx 3.0$.
   For our numerical comparison with the two--loop calculation of
Ref.\ \cite{Bellucci1} we made use of the parameters $L_i$ and $d_i$
corresponding to Tables 1 and 2 of Ref.\ \cite{Bellucci1}.
   In particular, from the numerical values of the parameters $a_1$, $a_2$
and $b$ of Table 2 of Ref.\ \cite{Bellucci1}
$$
 a_1^{BGS} = -39.0\,,\quad a_2^{BGS} = 12.5\,,\quad b^{BGS} = 3.0\,
$$
one obtains
\begin{equation}
d_1^{BGS} =-10.8\times 10^{-5}\,,\quad
d_2^{BGS} = 4.29\times 10^{-5}\,,\quad
d_3^{BGS} =-0.10\times 10^{-5}\,.
\label{diBGS}
\end{equation}

\begin{figure}
\epsfxsize=\textwidth
\epsfbox{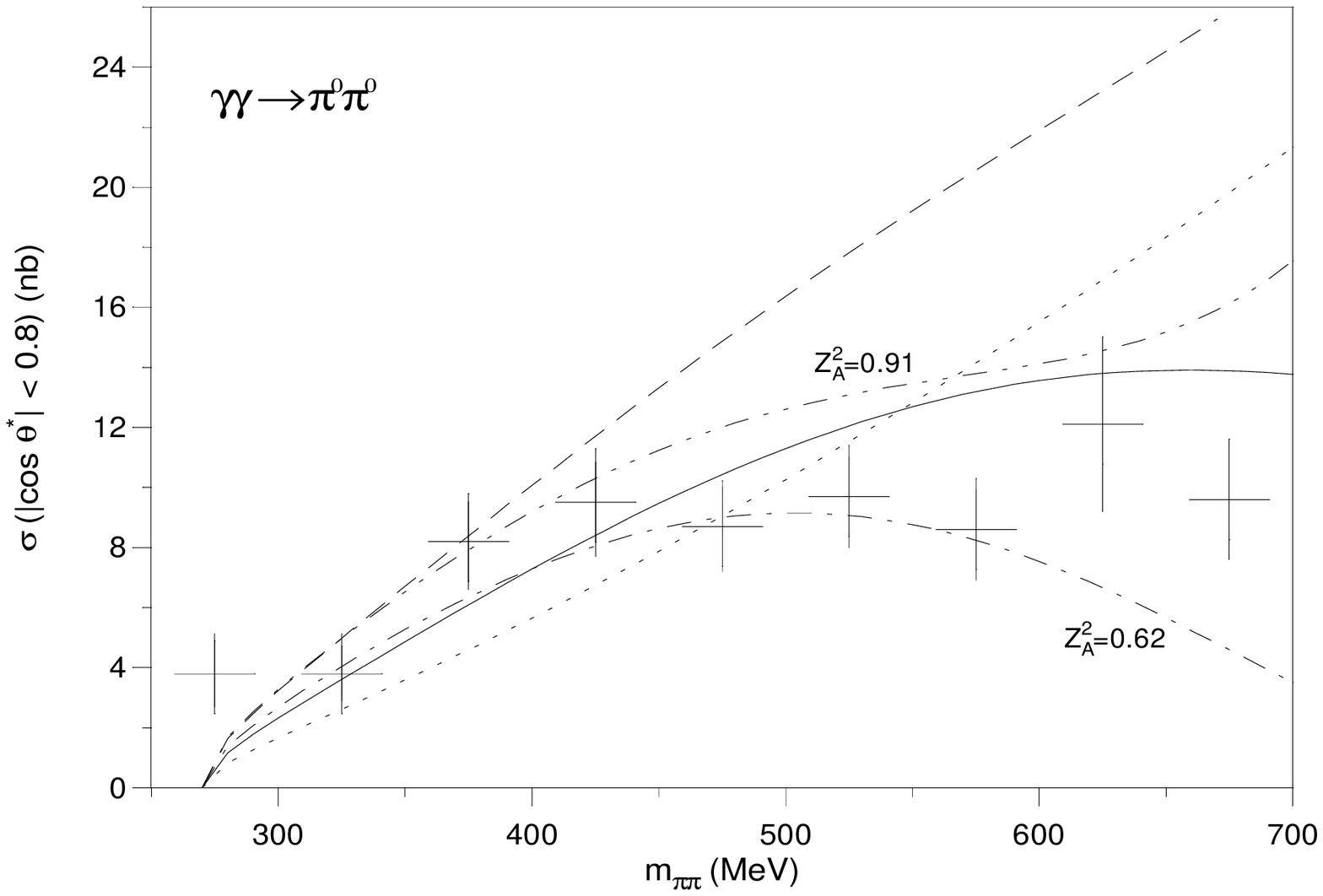}
\small
{\bf Fig.\ 1.}
   Cross section for $\gamma \gamma \to \pi^0 \pi^0$ as a function of the
invariant mass $W=m_{\pi^0\pi^0}$ for $W<0.7$ GeV and $|\cos\theta^*|
<0.8$ where $\theta^*$ is the angle between the beam axis and one
of the $\pi^0$ in the $\gamma\gamma$ center--of--mass system (c.m.s).
   The data are from the Crystal Ball experiment \cite{crystal-ball}.
   The dotted line represents the one--loop calculation at $O(p^4)$.
   The dashed line corresponds to the calculation at $O(p^6)$ without
reduction of the resonance degrees of freedom.
   The dash--dotted lines corresponding to two different values of the
parameter $Z_A^2$ are a measure for the uncertainty in the reduction
of the meson resonances.
   This  uncertainty is due to the difference between the NJL prediction and
the empirical value for the coupling constant $h_V$.
   The solid line corresponds to the values of the coefficients
$L_i$ and $d_i$ used in Ref.\ \cite{Bellucci1}.
\end{figure}

   Our predictions at $O(p^4)$ and $O(p^6)$ for the
$\gamma \gamma \to \pi^0 \pi^0$ cross section near threshold are shown in
Fig.\ 1.
   The calculation at $O(p^6)$ contains Born, one--loop and two--loop diagrams.
   In our two--loop calculation only diagrams which are factorizable and
which can be calculated analytically were taken into account.
   Two--loop box diagrams and acnode graphs cannot be calculated analytically
but the numerical estimates of Ref.\ \cite{Bellucci1} indicate that their
contributions are small.
    As was already discussed in Ref.\ \cite{Bellucci2}, the predictions of the
NJL model for the coefficients $d_1^{red}$ and $d_2^{red}$ are about a factor
one half smaller in comparison with the vector--meson--dominance
model (VMD) (see, Refs.\ \cite{Bijnens3,Bellucci2}).
   The coefficients $d_1$ and  $d_2$ in the VMD model can be obtained from
Eq.\ (\ref{dired}) by the replacement
\begin{equation}
  Z^4_A \quad \longrightarrow\quad \widetilde{Z}^4_A =
        \frac{6}{N_c}\bigg( \frac{16\pi h_V \mu}{m_V}\bigg)^2 = 0.82,
\end{equation}
   with $m_V = m_{\rho}$, and where the coupling constant
$h_V = 3.7\times 10^{-2}$ is extracted from the decays $V \to \pi \gamma$.
This has to be compared with the prediction of the NJL model,
$h^{NJL}_V=2.5\times 10^{-2}$ for $Z^2_A=0.62$.
   We have taken account of this uncertainty by showing the results for both
$Z^2_A=0.62$ and $\widetilde{Z}^2_A=0.91$.
   The results of our calculations with the parameters of
Ref.\ \cite{Bellucci1} are also shown in Fig.\ 1.
Numerically they are in a good agreement with Ref.\ \cite{Bellucci1};
even for $m_{\pi \pi}$ as large as 700 MeV the difference is only about
$7\%$.

\begin{figure}[t]
\begin{center}
Table 2.  Contribution of various diagrams to the $\eta\to\pi^0\gamma\gamma$
decay width.\\
$\Gamma^{exp}_{\eta\to\pi^0\gamma\gamma}=(0.84\pm0.18$)~eV.
\end{center}
\begin{center}
\renewcommand{\arraystretch}{1.4}
\begin{tabular}{|c|c|c|c|c|} \hline
\multicolumn{2}{|c|}{Amplitudes}  & Without
& \multicolumn{2}{|c|}{With reduction (eV)}\\ \cline{4-5}
\multicolumn{2}{|c|}{}            & reduction (eV)
& $Z_A^2=0.62$ & $Z_A^2=0.91$ \\ \hline
1-loop  & $\pi\pi$-loops          & $1.3 \cdot 10^{-3}$
& $1.3 \cdot 10^{-3}$ & $1.3 \cdot 10^{-3}$ \\ \cline{2-5}
$O(p^4)$ & $K  \overline{K}$-loops & $6.2 \cdot 10^{-3}$
& $6.2 \cdot 10^{-3}$ & $6.2 \cdot 10^{-3}$ \\ \hline
\multicolumn{2}{|c|}{Born $O(p^6)$}& 0.22
& 0.11 & 0.45 \\ \hline
1-loop  & $\pi\pi$-loops          & $1.9 \cdot 10^{-4}$
& $6.9 \cdot 10^{-5}$ & $8.6 \cdot 10^{-4}$ \\ \cline{2-5}
$O(p^6)$ & $K  \overline{K}$-loops & $4.1 \cdot 10^{-2}$
& $1.9 \cdot 10^{-3}$ & $2.7 \cdot 10^{-2}$ \\ \hline
2-loop  & $\pi\pi$-loops          & $3.2 \cdot 10^{-4}$
& $3.2 \cdot 10^{-4}$ & $3.2 \cdot 10^{-4}$ \\ \cline{2-5}
$O(p^6)$ & $\pi K $-loops          & $3.1 \cdot 10^{-3}$
& $3.1 \cdot 10^{-3}$ & $3.1 \cdot 10^{-3}$ \\ \cline{2-5}
        & $K  \overline{K}$-loops & $1.4 \cdot 10^{-5}$
& $1.4 \cdot 10^{-5}$ & $1.4 \cdot 10^{-5}$ \\ \hline
\multicolumn{2}{|c|}{Total}       & 0.14 & 0.11 & 0.35 \\ \hline
\end{tabular}
\renewcommand{\arraystretch}{1}
\end{center}
\end{figure}

   For the decay width of $\eta \to \pi^0 \gamma \gamma$ we obtain after
the reduction $0.11$ eV and $0.35$ eV corresponding to $Z^2_A=0.62$ and
$Z_A^2=0.91$, respectively.
   On the other hand, using the parameters of Eq.\ (\ref{diBGS}) one
finds $0.18$ eV.
   These results have to be compared with the experimental value
$(0.84\pm 0.18)$ eV \cite{PDG}.
   The contributions of different diagrams to the decay width are shown
in Table~2.
   These results clearly show the dominance of the Born contribution.
   It is a well--known fact that calculations of the decay width
at $O(p^6)$ tend to come out too small in comparison with the experimental
value \cite{Ametller,Bellucci2,Ko}.
   This failure indicates that either higher--order terms are required or
higher--order resonances have to be included or both.

\begin{figure}
\epsfxsize=\textwidth
\epsfbox{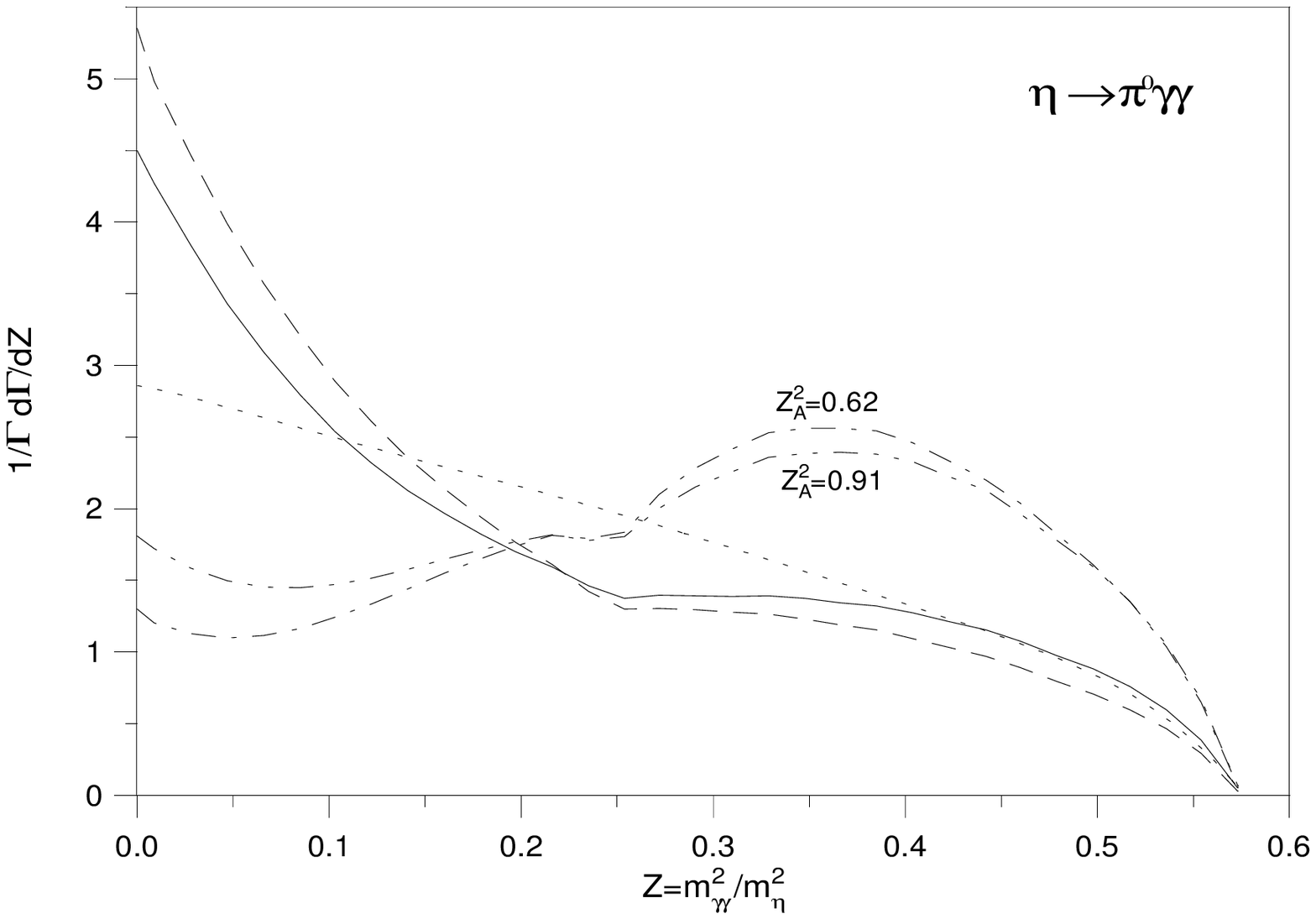}
\small
\noindent {\bf Fig.\ 2.}
   Normalized differential decay probability for
$\eta \to \pi^0 \gamma \gamma$ as a function of
$Z=m^2_{\gamma \gamma}/m^2_\eta$.
   The dotted line represents the phase space distribution.
   The dashed line corresponds to the calculation at $O(p^6)$ without
reduction of the resonances.
   The dash--dotted lines display the uncertainty in the
reduction of meson resonances for different values of the parameter
$Z_A^2$.
   The solid line corresponds to the values of the coefficients
$L_i$ and $d_i$ used in Ref.\ \cite{Bellucci1}.
\end{figure}

   Finally, we have also tried to fit the coefficients $d_1$, $d_2$ and
$d_3$.
   However, due to a strong correlation between the coefficients $d_1$ and
$d_3$ it was impossible to find a stable minimum from a fit
to the $\gamma \gamma \to \pi^0 \pi^0$ cross section and the
$\eta \to \pi^0 \gamma \gamma$ decay width.
   The strong correlation is related to the fact that the
$m_{\pi\pi}$ dependence of the $\gamma \gamma \to \pi^0 \pi^0$
cross section results from the interference between the Born amplitude
on the one hand and one-- and two--loop amplitudes on the other hand.
   Thus the experimental data are not sensitive enough to the
various Born contributions described by $d_i$.
   On the other hand, the Born contribution is dominating in the
$\eta \to \pi^0 \gamma \gamma$ decay.
   In Fig.\ 2 we show the normalized differential decay probability as
a function of $m^2_{\gamma \gamma}/m_\eta^2$.
   In this case the differential distribution is very sensitive to the
input parameters $d_i$.
   Thus data of the differential distribution would be of great
value for constraining these parameters.

   In conclusion, a self--consistent, quantitative description of
$\gamma\gamma\to\pi^0\pi^0$ {\em and} $\eta\to\pi^0\gamma\gamma$ data
at $O(p^6)$ is still problematic.
   A good description of the $\gamma\gamma\to\pi^0\pi^0$ cross section
has been achieved whereas a satisfactory, quantitative prediction of the
decay width seems to be beyond the reach of an ordinary calculation at
$O(p^6)$.


   The authors gratefully acknowledge fruitful discussions with
C.\ Bruno, G.\ Ecker, J.\ Gasser, E.\ A.\ Kuraev, H. Leutwyler,
M.\ R.\ Pennington, V.\ N.\ Pervushin, A.\ Schaale and M.\ K.\ Volkov.
   One of the authors (A.\ A.\ Bel'kov) would like to thank SFB 201 of
the Deutsche Forschungsgemeinschaft for its hospitality and financial
support during his stay at Mainz.
   This work was supported by the Russian Foundation for Fundamental
Research under grant No.~94-02-03973.



\begin{thebibliography}{99}
\bibitem{Terentev} M.\ V.\ Terent'ev, Sov.\ J.\ Nucl.\ Phys.\ {\bf 16}
                   (1973) 87.
\bibitem{Volkov1} M.\ K.\ Volkov and V.\ N.\ Pervushin,
                  Sov.\ J.\ Nucl.\ Phys.\ {\bf 22} (1976) 179.
\bibitem{polarizabilities} B.\ R.\ Holstein, Comments Nucl.\ Part.\ Phys.\
                           {\bf 19} (1990) 221;\\
                           D.\ Drechsel, L.\ V.\ Fil'kov, Z.\ Phys.\
                           {\bf A349} (1994) 177.
\bibitem{Belkov1} A.\ A.\ Bel'kov and V.\ N.\ Pervushin,
                  Sov.\ J.\ Nucl.\ Phys.\ {\bf 40} (1984) 616;\\
                  A.\ A.\ Bel'kov, E.\ A.\ Kuraev, and V.\ N.\ Pervushin,
                  Sov.\ J.\ Nucl.\ Phys.\ {\bf 40} (1984) 942.
\bibitem{crystal-ball} Crystal Ball Collaboration (H.\ Marsiske et al.),
                       Phys.\ Rev.\ {\bf D41} (1990) 3324.
\bibitem{Weinberg} S.\ Weinberg, Physica {\bf 96A} (1979) 327.
\bibitem{Gasser1} J.\ Gasser and H.\ Leutwyler,  Ann.\ Phys.\ {\bf 158}
                  (1984) 142; Nucl.\ Phys.\ {\bf B250} (1985) 465.
\bibitem{Bijnens1} J.\ Bijnens and F.\ Cornet, Nucl.\ Phys.\
                   {\bf B296} (1988) 557;\\
                   J.\ F.\ Donoghue, B.\ R.\ Holstein, and Y.\ C.\ Lin,
                   Phys.\ Rev.\ {\bf D37} (1988) 2423.
\bibitem{Bijnens3} J.\ Bijnens, S.\ Dawson, and G.\ Valencia,
                   Phys.\ Rev.\ {\bf D44} (1991) 3555.
\bibitem{dispersion} J.\ F.\ Donoghue and B.\ R.\ Holstein,
                     Phys.\ Rev.\ {\bf D48} (1993) 137;\\
                     D.\ Morgan and M.\ R.\ Pennington,
                     Phys.\ Lett.\ {\bf B272} (1991) 134;\\
                     M.\ R.\ Pennington, in {\it The DA$\Phi$NE Physics
                     Handbook}, Vol.\ II, p.\ 379 (Ed.\ L.\ Maiani,
                     C.\ Pancheri and N.\ Paver), INFN, Frascati, 1992.
\bibitem{Bellucci1} S.\ Bellucci, J.\ Gasser and M.\ E.\ Sainio,
                    Nucl.\ Phys.\ {\bf B423} (1994) 80;
                    ibid {\bf B431} (1994) 413 (Erratum).
\bibitem{Knecht} M.\ Knecht, B.\ Moussallam and J.\ Stern,
                 Nucl.\ Phys.\ {\bf B429} (1994) 125.
\bibitem{Ametller} Ll.\ Ametller, J.\ Bijnens, A.\ Bramon and F.\ Cornet,
                   Phys.\ Lett.\ {\bf B276} (1992) 185.
\bibitem{PDG} Particle Data Group, L.\ Montanet et al., Review of Particle
              Properties, Phys.\ Rev.\ {\bf D50} (1994) 1173.
\bibitem{Bellucci2}  S.\ Bellucci and C. Bruno, Report BUTP-94/26 (1994),
                     hep-ph/9502243.
\bibitem{Ko} P.\ Ko, Phys.\ Lett.\ {\bf B349} (1995) 555.
\bibitem{Bijnens2} J.\ Bijnens, C.\ Bruno and E.\ de Rafael,
                   Nucl.\ Phys.\ {\bf B390} (1993) 501.
\bibitem{Ng} J.\ N.\ Ng and D.\ J.\ Peters, Phys.\ Rev.\ {\bf D47} (1993)
             4939.
\bibitem{bosonization} M.\ K.\ Volkov, Ann.\ Phys.\ {\bf 157} (1984) 282;
                       Sov.\ J.\ Part.\ Nucl.\ {\bf 17} (1986) 186;
		        ibid.\ {\bf 24} (1993) 35;\\
                       D.\ Ebert and H.\ Reinhardt, Nucl.\ Phys.\ {\bf B271}
                       (1986) 188;\\
                       S.\ P.\ Klevansky, Rev.\ Mod.\ Phys.\ {\bf 64} (1992)
                       649;\\
                       A.\ A.\ Bel'kov et al., Int.\ J.\ Mod.\ Phys.\ {\bf C4}
                       (1993) 775;
                       Int.\ J.\ Mod.\ Phys.\ {\bf A8} (1993) 1313.
\bibitem{NJL} Y.\ Nambu and G.\ Jona-Lasinio,
              Phys.\ Rev.\ {\bf 122} (1961) 345; ibid.\ {\bf 124} (1961) 246.
\bibitem{Fearing} H.\ W.\ Fearing and S.\ Scherer,
                  TRIUMF report TRI-PP-94-68 (1994), hep-ph/9408346.
\bibitem{Volkov2} M.\ K.\ Volkov, Ann.\ Phys.\ {\bf 49} (1968) 202;
                  Fortschr.\ Phys.\ {\bf 22} (1974) 499.
\bibitem{propertime} J.\ Schwinger, Phys.\ Rev.\ {\bf 82} (1951) 664;\\
                     B.\ S.\ De Witt, {\it Dynamical theory of groups and
                     fields},
                     Gordon and Breach, New York (1965);\\
                     R.\ D. Ball, Phys. Rep. {\bf 182} (1989) 1.
\bibitem{Belkov2} A.\ A.\ Bel'kov, A.\ V.\ Lanyov and A.\ Schaale,
                  Dubna report JINR E2-95-238 (1995), hep-ph/9506237,
                  to appear in {\it Proc. IV Int. Workshop on Software
                  Engineering,
                  Artificial Intelligence for High Energy and Nuclear
                  Physics}, Pisa (Italy), Apr.\ 3--8, 1995, Ed.\ B.\ Denby.
\bibitem{Belkov3} A.\ A.\ Bel'kov, A.\ V.\ Lanyov, A.\ Schaale and S.\ Scherer,
                  Acta Physica Slovaca {\bf 45} (1995) 121.
\bibitem{reduction} G.\ Ecker, J.\ Gasser, A.\ Pich and E.\ de Rafael,
                    Nucl.\ Phys.\ {\bf B321} (1989) 311;\\
                    J.\ F.\ Donoghue, C.\ Ramirez, and G. Valencia,
                    Phys.\ Rev.\ {\bf D39} (1989) 1947.
\bibitem{Belkov4} A.\ A.\ Bel'kov, A.\ V.\ Lanyov and A.\ Schaale,
                  Acta Physica Slovaca {\bf 45} (1995) 135.
\bibitem{Reinhardt} H.\ Reinhardt and B.\ V.\ Dang, Nucl.\ Phys.\
                    {\bf A500} (1989) 563.
\end{thebibliography}
\end{document}